\documentclass[twocolumn,showpacs,aps,prl]{revtex4}
\usepackage{amssymb,times}
\usepackage{amsfonts}
\usepackage{dcolumn}
\usepackage{amsmath}
\usepackage{graphicx}
\usepackage{psfrag,pstricks}
\usepackage{amsmath}
\usepackage{amssymb}

\begin{document}

\title{Catch-Disperse-Release Readout for Superconducting Qubits}
\author{Eyob A. Sete$^1$, Andrei Galiautdinov$^{1,2}$, Eric Mlinar$^1$,
John M. Martinis$^3$, and Alexander N. Korotkov$^1$}
\affiliation{$^1$Department of Electrical Engineering, University of
California, Riverside, California 92521, USA \\
$^2$Department of Physics and Astronomy, University of Georgia, Athens,
Georgia 30602, USA\\
$^3$Department of Physics, University of California, Santa Barbara,
California 93106, USA}
\date{\today}

\begin{abstract}
We analyze a single-shot readout for superconducting qubits via
the controlled catch, dispersion, and release of a microwave field. A
tunable coupler is used to decouple the microwave resonator from the
transmission line during the dispersive qubit-resonator interaction,
thus circumventing damping from the Purcell effect. We show that if
the qubit frequency tuning is sufficiently adiabatic, a fast
high-fidelity qubit readout is possible even in the strongly
nonlinear dispersive regime. Interestingly, the Jaynes-Cummings
nonlinearity leads to the quadrature squeezing of the resonator
field below the standard quantum limit, resulting in a significant
decrease of the measurement error.
\end{abstract}
\maketitle

A fast high-fidelity qubit readout plays an important role in quantum
information processing. For superconducting qubits various nonlinear
processes have been used to realize a single-shot readout
\cite{Cooper-04,Ast04,Sid06,Lupascu-06,Mal09,Reed-10}. The linear
dispersive readout in the circuit quantum electrodynamics (cQED)
setup \cite{Wal05,Blais-04} became sufficiently sensitive for the
single-shot qubit measurement only recently \cite{Sid12,Ris12}, with
development of near-quantum-limited superconducting parametric
amplifiers \cite{Sid12,Ris12,Bergeal-10}. In particular, readout
fidelity of $94\%$ for flux qubits \cite{Sid12} and $97\%$ for
transmon qubits \cite{Ris12} has been realized (see also
\cite{Devoret-13}). With increasing coherence time of
superconducting qubits into 10-100 $\mu$s range \cite{Byl11,Pai11},
fast high-fidelity readout becomes practically important, for
example, for reaching the threshold of quantum error correction
codes \cite{Fowler-12}, for which the desired readout time is less
than 100 ns, with fidelity above 99\%.

A significant source of error in the currently available cQED
readout schemes is the Purcell effect \cite{Pur46} --- the
cavity-induced relaxation of the qubit due to the always-on coupling
between the resonator and the outgoing transmission line. The
Purcell effect can be reduced by increasing the qubit-resonator
detuning; however, this reduces the dispersive interaction and
increases measurement time. Several proposals to overcome the
Purcell effect have been put forward, including the use of the
Purcell filter \cite{Red10} and the use of a Purcell-protected qubit
\cite{Bla11}. Here we propose and analyze a cQED scheme which avoids
the Purcell effect altogether by decoupling the resonator from the
transmission line during the dispersive qubit-resonator interaction.

Similar to the standard cQED measurement
\cite{Wal05,Blais-04,Sid12,Ris12}, in our method (Fig.\ 1) the qubit
state affects the dispersive shift of the resonator frequency, that
in turn changes the phase of the microwave field in the resonator,
which is then measured via homodyne detection. However, instead of
measuring continuously, we perform a sequence of three operations:
``catch'', ``disperse'', and ``release'' of the microwave field.
During the first two stages a tunable coupler decouples the outgoing
transmission line from the resonator (we assume using the coupler
recently realized in \cite{Mar12}; see also \cite{Mar11}). This
automatically eliminates the problems associated with the Purcell
effect, as coupling to the incoming microwave line can be made very
small \cite{Mar12}.

\begin{figure}[t]
\includegraphics[width=6.4cm]{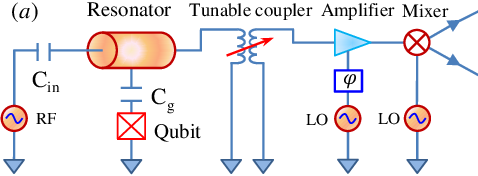}
\includegraphics[width=6.4cm]{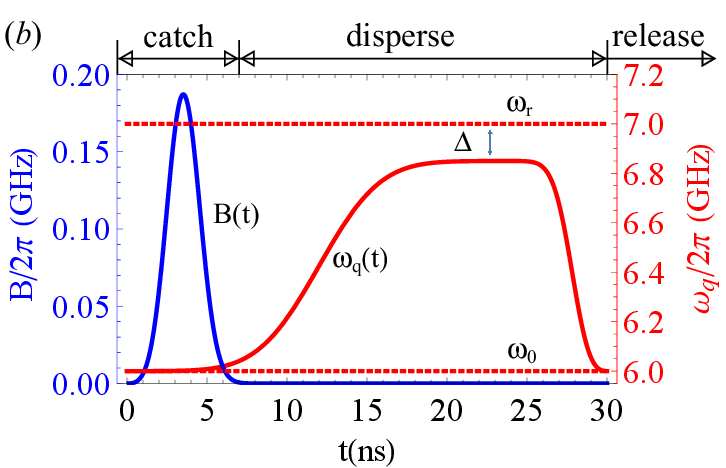}
  \caption{(a) Schematic of the measurement setup.
The radio frequency (RF) source produces a microwave pulse, which
populates the resonator via a small capacitor $C_{\text{in}}$. The
resonator photons then interacts with a capacitively
($C_{\text{g}}$) coupled qubit. The interaction with the outgoing
transmission line is controlled by a tunable coupler, which releases
photons at the end of the procedure. The released field is then
amplified and mixed with the local oscillator (LO) signal to be
measured via homodyne detection.
 (b) The RF pulse $B(t)$ (blue curve) and varying qubit frequency
$\omega_{\rm q}(t)$ (red curve), with approximate indication of the
``catch'', ``disperse'', and ``release'' stages. Dashed lines show
the resonator frequency $\omega_{\rm r}$ and initial/final qubit
frequency $\omega_0$; $\Delta=\omega_{\rm r}-\omega_{\rm q}$ is the
detuning at the ``disperse'' stage.  {\bf }
 }\label{fig1}
\end{figure}

\begin{figure}[t]
\includegraphics[width=4.5cm]{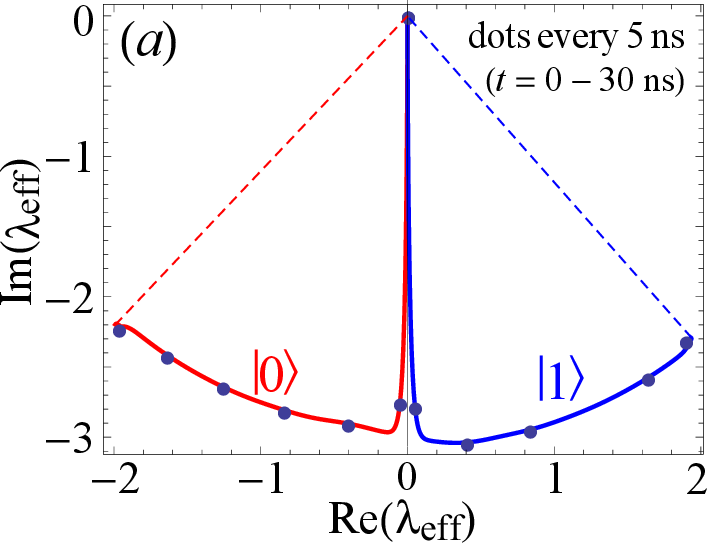}~~\includegraphics[width=4.3cm]{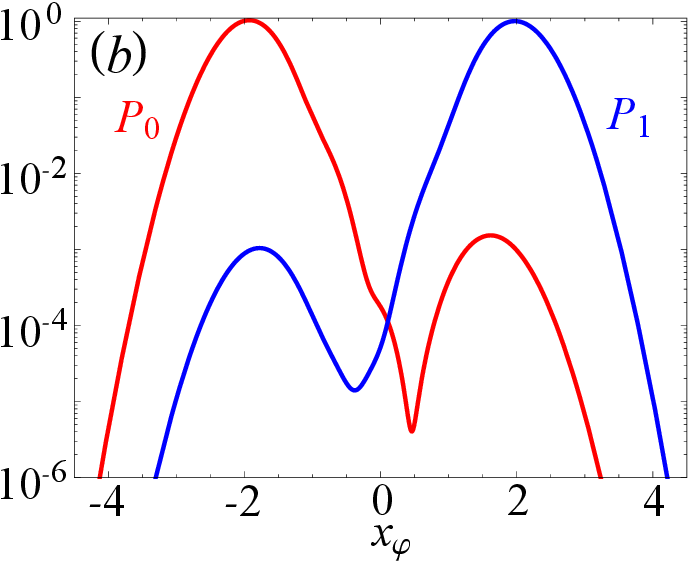}
\caption{(a) Evolution in time of the effective
field amplitude $\lambda_{\text{eff}}$ on the phase plane
for initial qubit states $|0\rangle$ and $|1\rangle$, computed
numerically. The
dots indicate time moments $t=0$, 5, 10, 15, 20, 25, and 30 ns. (b)
Corresponding probability distributions $P_0(x_\varphi)$ and
$P_1(x_\varphi)$ for measurement (at $t=t_{\rm f}$) of the optimum
quadrature $x_{\varphi}$. Side bumps of $P_0$ and $P_1$ are due to
non-adiabaticity. We used $\Delta/2\pi=50$
MHz, $|\lambda_{\rm in}|^2=9$, $\sigma_{\rm q}=3$ ns, $t_{\rm q}=3.25$ ns, $t_{\rm
qe}=30$ ns, and $t_{\rm f}=32$ ns.
    } \label{fig2} \end{figure}

  During the ``catch'' stage, the initially empty resonator is driven by a
microwave pulse and populated with $\sim$10 photons. At this stage
the qubit is far detuned from the resonator [Fig.\ 1(b)], which
makes the dispersive coupling negligible and allows the creation of
an almost-perfect coherent state in the resonator.
  At the next ``disperse'' stage of the measurement, the qubit
frequency is adiabatically tuned closer to the resonator frequency
to produce a strong qubit-resonator interaction (it may even be
pushed into the nonlinear regime). During this interaction, the
resonator field amplitudes ($\lambda_{\rm eff}$) associated with the
initial qubit states $|0\rangle$ and $|1\rangle$ rapidly accumulate
additional phases and separate in the complex phase plane [see Fig.\
\ref{fig2}(a)]. Finally, at the last ``release'' stage of the
measurement, after the qubit frequency is again detuned from the
resonator, the resonator photons are released into the outgoing
transmission line. The signal is subsequently amplified (by a
phase-sensitive parametric amplifier) and sent to the mixer where
the homodyne detection is performed.

    With realistic parameters for superconducting
qubit technology, we numerically show that the measurement of 30--40
ns duration can be realized with an error below $10^{-3}$,
neglecting the intrinsic qubit decoherence. The latter assumption
requires the qubit coherence time to be over 40 $\mu$s, which is
already possible experimentally \cite{Pai11}. It is interesting that
because of the interaction nonlinearity \cite{Bishop-10,Bla10},
increasing the microwave field beyond $\sim$10 photons only slightly
reduces the measurement time. The nonlinearity also gives rise to
about $\sim$50\% squeezing of the microwave field (see
\cite{squeezing-exp,squeezing-Wilhelm}), which provides an
order-of-magnitude reduction of the measurement error.

We consider a superconducting phase or transmon
qubit capacitively coupled to a microwave resonator [Fig.\
\ref{fig1}(a)]. For simplicity we start with considering a two-level
qubit (the third level will be included later) and describe the
system by the Jaynes-Cummings (JC) Hamiltonian \cite{Blais-04} with
a microwave drive ($\hbar=1$)
\begin{align}\label{H}
  H&=\omega_{\rm q}(t) \sigma_{+}\sigma_{-} +\omega_{\rm r}a^{\dag} a
+ \textit{g} (a \sigma_{+} + \sigma_{-}a^{\dag})
     \notag\\
  &+ B(t)a^{\dag}e^{-i\omega t}+B^{*}(t) a e^{i\omega t},
\end{align}
where $\omega_{\rm q}(t)$ and $\omega_{\rm r}$ are, respectively,
the qubit and the resonator frequencies, $\sigma_{\pm}$ are the
rasing and lowering operators for the qubit, $a$ ($a^{\dag}$) is the
annihilation (creation) operator for the resonator photons,
$\textit{g}$ (assumed real) is the qubit-resonator coupling, $B(t)$
and $\omega$ are the effective amplitude and the frequency of the
microwave drive, respectively. In this work we assume $\omega =
\omega_{\rm r}$.

    For the microwave drive $B(t)$ and the qubit frequency $\omega_q(t)$
[Fig.\ 1(b)] we use Gaussian-smoothed step-functions:
$B(t)=0.5B_{0}\{\text{Erf}[(t-t_{\rm B})/\sqrt{2}\sigma_{\rm B}]
-\text{Erf}[(t-t_{\rm B}-\tau_{\rm B})/\sqrt{2}\sigma_{\text{B}}]\}$
and $\omega_{\text{q}}(t)=\omega_{0}+0.5(\Delta_{0}-\Delta)
\{\text{Erf}[(t-t_{\rm
q})/\sqrt{2}\sigma_{\text{q}}]-\text{Erf}[(t-t_{\rm
qe})/\sqrt{2}\sigma_{\text{qe}}]\}$, where $t_{\rm B}$, $t_{\rm
B}+\tau_{\rm B}$, $t_{\rm q}$, and $t_{\rm qe}$ are the centers of
the front/end ramps, and $\sigma_{\rm B}$, $\sigma_{\rm q}$, and
$\sigma_{\rm qe}$ are the corresponding standard deviations. In
numerical simulations we use $\sigma_{\rm B}=\sigma_{\rm qe}=1$ ns
(typical experimental value for a short ramp) while we use longer
$\sigma_{\rm q}$ to make the qubit front ramp more adiabatic. Other
fixed parameters are: $\textit{g}/2\pi=30$ MHz, $\tau_{\rm B}=1$ ns,
$t_{\rm B}=3$ ns, $\omega_{\rm r}/2\pi=7$ GHz, and
$\omega_{0}/2\pi=6$ GHz, so that initial/final detuning
$\Delta_0=\omega_{\rm r}-\omega_0$ is 1 GHz, while the
disperse-stage detuning $\Delta$ is varied. The measurement starts
at $t=0$ and ends at $t_{\rm f}=t_{\rm qe}+2\sigma_{\rm qe}$ when
the field is quickly released \cite{note-shape}.

Let us first consider a simple
dispersive scenario at large qubit-resonator detuning, $|\Delta |
\gg \textit{g}\sqrt{\overline{n}+1}$, where $\overline{n}$ is the
average number of photons in the resonator. In this case, the system
is described by the usual dispersive Hamiltonian \cite{Blais-04} $
H_{d}=(\omega_{0}-\textit{g}^2/\Delta)\sigma_{z}/2
+(\omega_{r}-\sigma_{z}\textit{g}^2/\Delta)a^{\dag}a$, where
$\sigma_z$ is the Pauli matrix. After the short ``catch'' stage the
system is in a product state
$(\alpha|0\rangle+\beta|1\rangle)|\lambda_{{\rm in}}\rangle$, where
$\alpha$ and $\beta$ are the initial qubit state amplitudes and
$\lambda_{{\rm in}}$ is the amplitude of the coherent resonator
field, $\lambda_{\rm in}=-i\int B(t)\, dt$ (so
$\overline{n}=|\lambda_{\rm in}|^2$).
   Then during the ``disperse'' stage the qubit-resonator state
becomes entangled,
$\alpha|0\rangle|\lambda_{0}(t)\rangle+\beta|1\rangle|\lambda_{1}(t)\rangle$,
with  $\lambda_{0}=\lambda_{\rm in}e^{-i\phi},
\lambda_{1}=\lambda_{\rm in}e^{i\phi}$, and
$\phi(t)=\int_{0}^{t}[\textit{g}^2/\Delta(t')]dt'$.

The distinguishablity of the two resonator states depends on their
separation $|\delta\lambda| \equiv
|\lambda_{1}-\lambda_{0}|=2|\lambda_{\rm in}|\sin|\phi|$ (see
numerical results in Fig.\ \ref{fig2}). The released coherent states
are measured via the homodyne detection using the optimal quadrature
connecting $\lambda_{0}$ and $\lambda_{1}$, i.e.,\ corresponding to
the angle $\varphi=\text{arg}(\lambda_{1}-\lambda_{0})$.
   We rescale the measurement results to the dimensionless field
quadrature ${\hat{x}}_{\varphi}=(ae^{-i\varphi} +
a^{\dag}e^{i\varphi})/2$, which corresponds to the $\varphi$-angle
axis in the phase space of Fig.\ \ref{fig2}(a).
 In resolving the two coherent states, we are essentially
distinguishing two Gaussian probability distributions,
$P_{0}({x_\varphi})$ and $P_{1}({x_\varphi})$, centered at
$\pm|\delta\lambda|\sigma_{\text{coh}}$ with
$\sigma_{\text{coh}}=1/2$ being the coherent-state width (standard
deviation) for both distributions. Then the measurement error has a
simple form
\begin{equation}\label{Error}
E=\frac{1}{2}\int_{-\infty}^{\infty}
\text{min}(P_{0},P_{1})\,d{x_\varphi} =\frac{1 -\text{Erf}
(|\delta\lambda| \sqrt{\eta/2})}{2} ,
\end{equation}
where $\eta=\eta_{\text{\rm col}}\eta_{\text{\rm amp}}$ is the
detection efficiency \cite{Kor11}, which includes the collection
efficiency $\eta_{\text{col}}$ and quantum efficiency of the
amplifier $\eta_{\text{amp}}$. Unless mentioned otherwise, we assume
$\eta=1$, which corresponds to a quantum-limited phase-sensitive
amplifier (for a phase-preserving amplifier $\eta\leq 1/2$).

In general the JC qubit-resonator interaction (\ref{H}) is
non-linear for $|\lambda_{\rm in}|^2\agt \overline{n}_{\rm
crit}\equiv \Delta^2/4g^2$ \cite{Blais-04} and the resonator states
are not coherent. The measurement error $E$ is still given by the
first part of Eq.\ (\ref{Error}), while the probability
distributions $P_{0,1}(x_\varphi)$ of the measurement result for the
qubit starting in either state $|0\rangle$ or $|1\rangle$ can be
calculated in the following way. Assuming an instantaneous release
of the field, we are essentially measuring the operator
${\hat{x}}_{\varphi}$. Therefore the probability $P(x_\varphi)$ for
the ideal detection ($\eta=1$) can be calculated by converting the
Fock-space density matrix $\rho_{nm}$ describing the resonator
field, into the ${{x}}_{\varphi}$-basis,
  thus obtaining
$P({x_\varphi})=\sum_{nm}\psi_{n}(x_\varphi
)\rho_{nm}(t)\psi^{*}_m(x_\varphi) e^{-i(n-m)\varphi}$, where
$\psi_n(x)$ is the standard $n$th-level wave function of a harmonic
oscillator. For a non-instantaneous release of the microwave field
the calculation of $P({x_\varphi})$ is non-trivial; however, since
the qubit is already essentially decoupled from the resonator, the
above result for $P({x_\varphi})$ remains the same \cite{W-M-book}
for optimal time-weighting of the signal. In the case of a non-ideal
detection ($\eta<1$) we should take a convolution of the ideal
$P(x_\varphi)$ with the Gaussian of width
$\sqrt{\eta^{-1}-1}\,\sigma_{\text{coh}}$.
  Calculation of the optimum phase angle $\varphi$ minimizing the error
is non-trivial in the general case. For simplicity we still use the
natural choice
$\varphi=\text{arg}(\lambda_{\text{eff},1}-\lambda_{\text{eff},0})$,
where the effective amplitude of the resonator field \cite{Ger05} is
defined by $\lambda_{\text{eff}}=\sum_n \sqrt{n} {\rho}_{n, n-1}$.
The field density matrix ${\rho}_{n m}$ is calculated numerically
using the Hamiltonian (\ref{H}) and then tracing over the qubit.

    Extensive numerical simulations allowed us to identify two
main contributions to the measurement error $E$ in our scheme. The
first contribution is due to the insufficient separation of the
final resonator states $|\lambda_{\rm eff,1}\rangle$ and
$|\lambda_{\rm eff,0}\rangle$, as described above. However, there
are two important differences from the simplified analysis: the JC
nonlinearity may dramatically change $|\delta \lambda|$ and it also
produces a self-developing squeezing of the resonator states in the
quadrature $x_\varphi$, significantly decreasing the error compared
with Eq.\ (\ref{Error}) (both effects are discussed in more detail
later). The second contribution to the measurement error is due to
the nonadiabaticity of the front ramp of the qubit frequency pulse
$\omega_{q}(t)$, which leads to the population of ``wrong'' levels
in the eigenbasis. This gives rise to the side peaks (``bumps'') in
the probability distributions $P_{0,1}(x_\varphi)$, as can be seen
in Fig.\ \ref{fig2}(b) (notice their similarity to the experimental
results \cite{Sid12,Ris12}, though the mechanism is different).
During the dispersion stage these bumps move in the ``wrong''
direction, halting the exponential decrease in the error, and thus
causing the error to saturate. The nonadiabaticity at the rear ramp
of $\omega_{q}(t)$ is not important because the moving bumps do not
have enough time to develop. Therefore the rear ramp can be steep,
while the front ramp should be sufficiently smooth [Fig.\ 1(a)] to
minimize the error.

\begin{figure}[t]
\includegraphics[width=4.4cm]{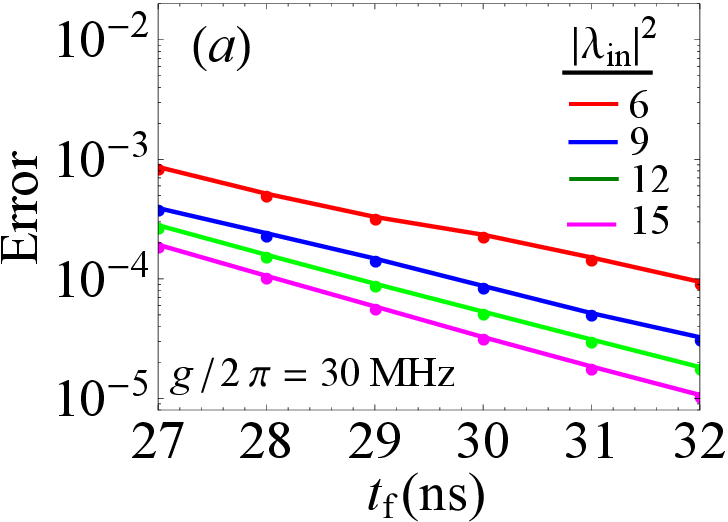}\includegraphics[width=4.4cm]{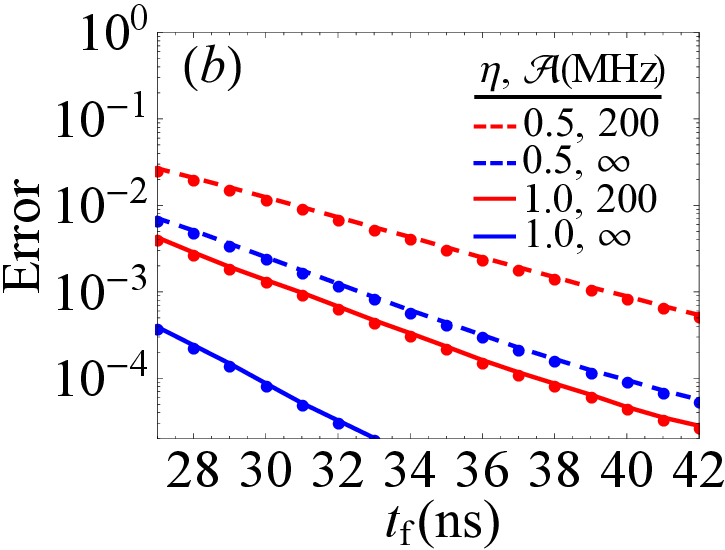}
\caption{Optimized measurement error $E$ vs measurement time $t_{\rm
f}$ (optimization is over $\Delta$, $\sigma_{\rm q}$, and $t_{\rm
q}$). (a) For two-level qubit and for mean photon number
$|\lambda_{\rm in}|^2=6, 9, 12$, and $15$. (b) For $|\lambda_{\rm
in}|^2=9$ and $\eta=1$ or $1/2$ (e.g.\ for a phase-preserving
amplifier),  taking into account the qubit level $|2\rangle$ (with
anharmonicity ${\mathcal A}/2\pi=200$ MHz) or assuming a two-level
qubit ($\mathcal{A}=\infty$).}\label{fig3}
\end{figure}

    Now, let us discuss the effect of nonlinearity
(when $|\lambda_{\rm in}|^2 > \overline{n}_{\rm crit}$) on the
evolution of $\lambda_{\rm eff,0}$ and $\lambda_{\rm eff,1}$ during
the disperse stage. Since the RF drive is turned off, the
interaction described by the Hamiltonian \eqref{H} occurs only
between the pairs of states $|0,n\rangle$ and $|1,n-1\rangle$ of the
JC ladder. Therefore, if the front ramp of the qubit frequency pulse is
adiabatic, the pairs of the JC {\it eigenstates} evolve only by
accumulating their respective phases while maintaining their
populations.  Then for the qubit initial state $|0\rangle$, the
qubit-resonator wavefunction evolves approximately as
$|\psi_{0}(t)\rangle \simeq e^{-|\lambda_{\rm in}|^2/2} \sum_{n}
(\lambda_{\rm in}^n/\sqrt{n!}) e^{-i\phi_{0,n}(t)}
|\overline{0,n}\rangle$, where the overbar denotes the (dressed)
eigenstate and $\phi_{0,n}(t)=\int_{t_{\rm
D}}^{t}dt'[\sqrt{\Delta(t')^2+4\textit{g}^2n} - \Delta(t')]/2$ is
the accumulated phase, with $t_{\rm D}=t_{\rm B}+\tau_{\rm B}/2$
being the center of the $B(t)$-pulse, which is crudely the start of
the dispersion. Similarly, if the qubit starts in state $|1\rangle$
(following the ideology of Ref.\ \cite{RezQu}, we then use
$\overline{|10\rangle}$ as the initial state), the state evolves as
$|\psi_{1}(t)\rangle \simeq e^{-|\lambda_{\rm in}|^2/2}
\sum_{n}(\lambda_{\rm in}^n/\sqrt{n!}) e^{i\phi_{1,n}(t)}
|\overline{1,n}\rangle$, where $\phi_{1,n}(t)=\int_{t_{\rm
D}}^{t}dt' [\sqrt{\Delta(t')^2+4\textit{g}^2(n+1)}-\Delta(t')]/2$.
Using the above definition of $\lambda_{\rm eff}$ and assuming
$|\lambda_{\rm in}|^2\gg 1$ we derive an approximate formula
    \begin{equation}\label{L0}
  \lambda_{\text{eff},0}=\lambda_{\rm in}
 \exp\left[-i\int_{t_{\rm D}}^t
 \frac{\textit{g}^2}{\sqrt{\Delta(t')^2+4\textit{g}^2
 |\lambda_{\rm in}|^2}} \, dt'\right].
    \end{equation}
The corresponding expression for $\lambda_{\text{eff},1}$ can be
obtained by replacing $-i$ with $i$ and $|\lambda_{\rm in}|^2$ with
$|\lambda_{\rm in}|^2+1$. These formulas agree well with our
numerical results.

   Equation \eqref{L0} shows that a decrease in detuning
leads to an increase in the rotation speed of
$\lambda_{\text{eff}}$. However, in the strongly nonlinear regime
$|\lambda_{\rm in}|^2\gg \overline{n}_{\rm crit}$, the angular speed
saturates at
$d(\text{arg}(\lambda_{\text{eff,0/1}}))/dt=\mp\textit{g}/2|\lambda_{\rm
in}|$. Thus, the rate at which the $\lambda_{\text{eff,1}}$ and
$\lambda_{\text{eff,0}}$ separate is limited by
    \begin{equation}
d|\delta\lambda|/dt\leq |\textit{g}|,
    \label{rate}\end{equation}
which does not depend on $|\lambda_{\rm in}|$. This means that
the measurement time should not improve much with increasing the
mean number of photons $|\lambda_{\rm in}|^2$ in the resonator,
as long as it is sufficient for distinguishing the states with a
desired fidelity (crudely, $|\lambda_{\rm in}|^2\agt 7/\eta$
for $E\alt 10^{-4}$).


Figure \ref{fig3}(a) shows the results of a three-parameter
optimization of the measurement error $E$ for several values of the
average number of photons in the resonator, $|\lambda_{\rm in}|^2$
(assuming $\eta=1$). The optimization parameters are the
qubit-resonator detuning $\Delta$, the width $\sigma_{q}$, and the
center $t_{q}$ of the qubit front ramp. We see that for nine photons in
the resonator the error of $10^{-4}$ can be achieved with 30 ns
measurement duration, excluding time to release and measure the
field. The optimum parameters in this case are: $\Delta/2\pi=60$
MHz, $\sigma_q=4.20$ ns, and $t_q=3.25$ ns (this is a strongly
nonlinear regime: $|\lambda_{\rm in}|^2/\overline{n}_{\rm crit}=9$).
As expected from the above discussion, increasing the mean photon
number to 12 and 15 shortens the measurement time only slightly (by
1 ns and 2 ns, keeping the same error). The dashed blue curve in
Fig.\ \ref{fig3}(b) shows the optimized error for $|\lambda_{\rm
in}|^2=9$ and imperfect quantum efficiency $\eta=1/2$. As we see,
the measurement time for the error level of $10^{-4}$ increases to
40 ns, while the error of $10^{-3}$ is achieved at $t_{\rm f}=32$
ns.

    So far, we considered the two-level model for the qubit.
However, real superconducting qubits are only slightly anharmonic
oscillators, so the effect of the next excited level $|2\rangle$ is
often important. It is straightforward to include the level
$|2\rangle$ into the Hamiltonian (\ref{H}) by replacing its first
term with $\omega_{\rm q}|1\rangle\langle 1|+ (2\omega _{\rm
q}-{\mathcal A})|2\rangle\langle 2|$, where ${\mathcal A}$ is the
anharmonicity. The dispersion can then be understood as due to
repulsion of three eigenstates: $\overline{|0,n\rangle}$,
$\overline{|1,n-1\rangle}$, and $\overline{|2,n-2\rangle}$. As the
result, $\lambda_{\rm eff,0}$ rotates on the phase plane faster than
in the two-level approximation, while $\lambda_{\rm eff,1}$ rotates
slower (sometimes even in the opposite direction). The Supplemental
Material \cite{SM} illustrates  evolution of the resonator Wigner
function corresponding to initial qubit states $|0\rangle$ and
$|1\rangle$. In Fig.\ \ref{fig3}(b), we present the optimized error
for ${\mathcal A}/2\pi =200$ MHz (a typical value for transmon and
phase qubits), $|\lambda_{\rm in}|^2=9$ and $\eta=1$ (solid-red
curve) or $\eta=1/2$ (dashed-red curve). An error of $10^{-3}$ can
be achieved with 31 ns ($\eta=1$) and 39 ns ($\eta=1/2$) measurement
durations.

\begin{figure}[t]
\includegraphics[width=4.4cm]{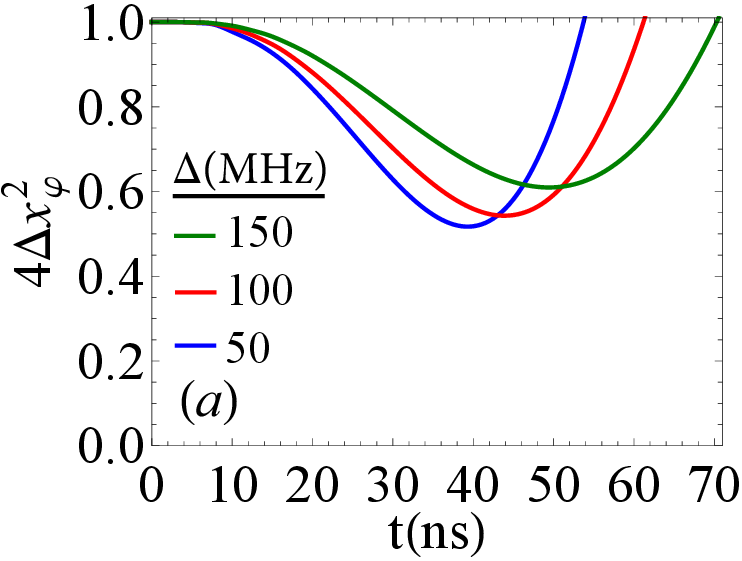}\includegraphics[width=4.4cm]{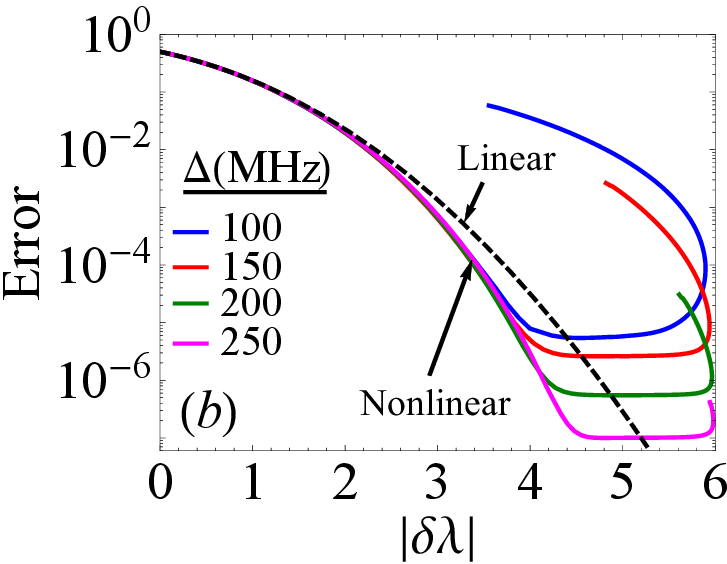}
\caption{(a) Evolution of the quadrature squeezing (the qubit is
initially in state $|0\rangle$). (b) Measurement error vs
$|\delta\lambda|$ calculated numerically in the nonlinear regime
(solid lines) and using the linear approximation \eqref{Error}
(dashed line); here the evolution stops at 98 ns. $|\lambda_{\rm
in}|^2=9$, $\sigma_{q}=4\text{ns}$, $t_{q}=3.25$ ns.
 }  \label{squz}
\end{figure}

We next discuss the self-generated quadrature squeezing of the
microwave field induced by the JC nonlinearity. To quantity the
degree of squeezing, we calculate the variance $ \Delta
x_{\varphi}^2=\langle x_{\varphi}^2\rangle - \langle
x_{\varphi}\rangle^2=1/4+ \langle a^{\dag}a\rangle/2 -|\langle
a\rangle|^2/2+{\rm Re}[(\langle a^2\rangle-\langle
a\rangle^2)e^{-2i\varphi}]/2$. For a coherent field $\Delta
x_{\varphi}^2=1/4$, thus the state is squeezed \cite{Ger05} when
$4\Delta x_{\varphi}^2<1$. Figure \ref{squz}(a) shows evolution of
$4\Delta x_{\varphi}^2$ when the initial qubit state is $|0\rangle$,
for $\eta=1$ and assuming a two-level qubit (a similar result is
obtained for qubit initially in state $|1\rangle$). Notice that at
first the field stays coherent, which is due to the linearity of the
qubit-resonator interaction at large detuning. Later on, however,
the interaction becomes nonlinear due to decreased detuning and
leads to quadrature squeezing reaching the level of $\sim$50\% for
$\Delta/2\pi \alt 100$ MHz (see \cite{SM} for the Wigner function
evolution). Figure \ref{squz}(b) shows the measurement error as a
function of $|\delta\lambda|$ in the nonlinear regime calculated
numerically (solid curves) and in the linear regime based on Eq.\
\eqref{Error} (dashed curve). As expected, with the squeezing
developing, the error becomes significantly smaller than the
linear (analytical) prediction, for instance, up to a factor of 30 for
$\Delta/2\pi=250$ MHz. Note also that the error shown in Fig.\
\ref{squz}(b) saturates in spite of increasing separation
$|\delta\lambda|$. This is because of the nonadiabatic error
discussed above.

    We do not focus on the quantum nondemolition
(QND) \cite{Braginsky-Khalili} property of the readout, because in
the proposed implementation of the surface code \cite{Fowler-12} the
measured qubits are reset, so the QNDness is not important. For the
results presented in Fig.\ \ref{fig3} the non-QND-ness (probability
that the initial states $|00\rangle$ and $\overline{|10\rangle}$ are
changed after the procedure) is crudely about 5\%, which is mainly
due to nonadiabaticity of the rear ramp. It is possible to strongly
decrease the non-QND-ness by using smoother rear ramp, but  it cannot
be reduced below a few times $(\textit{g}/\Delta_0)^2$, essentially
because of the Purcell effect during the release stage. Furthermore,
we do not consider the measurement-induced dephasing of the qubit,
since our readout is not intended for a continuous qubit monitoring
or a quantum feedback. We neglect the qubit relaxation and
excitation due to ``dressed dephasing'' \cite{dressed-deph} because
its rate is smaller than the intrinsic pure dephasing, which for
transmons is usually smaller than intrinsic relaxation.

In conclusion, we analyzed a fast high-fidelity readout for
superconducting qubits in a cQED architecture using the controlled
catch, dispersion, and release of the microwave photons. This
readout uses a tunable coupler to decouple the resonator from the
transmission line during the dispersion stage of the measurement,
thus avoiding the Purcell effect.  Our approach may also be used as
a new tool to beat the standard quantum limit via self-developing
field squeezing, directly measurable using the state-of-the-art
parametric amplifiers.

The authors thank Farid Khalili, Konstantin Likharev, and Gerard
Milburn for useful discussions. This research was funded by the
Office of the Director of National Intelligence (ODNI), Intelligence
Advanced Research Projects Activity (IARPA), through the Army
Research Office Grant No. W911NF-10-1-0334. All statements of fact,
opinion or conclusions contained herein are those of the authors and
should not be construed as representing the official views or
policies of IARPA, the ODNI, or the U.S. Government. We also
acknowledge support from the ARO MURI Grant No. W911NF-11-1-0268.


\begin{thebibliography}{99}
\bibitem{note-shape} Actually, for $\omega_{\rm q}(t)$ we
additionally use small compensating ramps at the beginning and end
of the procedure to provide the exact value $\omega_0$ and to zero
$\dot\omega_{\rm q}(t)$ at $t=0$ and $t= t_{\rm f}$.
\end{thebibliography}
\end{document}